\RequirePackage{fix-cm}
\documentclass[twocolumn,epjc3]{svjour3}  
\smartqed  
\RequirePackage{graphicx}

\usepackage{amssymb}
\usepackage{amsmath}
\usepackage{color}
\usepackage[colorlinks]{hyperref}

\def\w2{\tilde w^2}
\def\ws2{1}

\newlength{\lslash}

\journalname{Eur. Phys. J. C}

\begin{document}
\title{A generalized Higgs potential \\ with two degenerate minima \\ for a dark QCD matter scenario}
\author{
 Matthias F.M. Lutz\thanksref{e1,addr1,addr3}
\and Yonggoo Heo\thanksref{addr2}  
\and Xiao-Yu Guo\thanksref{addr1}
}
\thankstext{e1}{e-mail: m.lutz@gsi.de}

\institute{GSI Helmholtzzentrum f\"ur Schwerionenforschung GmbH,
Planck Str. 1, 64291 Darmstadt, Germany \label{addr1}
\and
 Suranaree University of Technology, Nakhon Ratchasima, 30000, Thailand \label{addr2}
\and 
Technische Universit\"at Darmstadt, D-64289 Darmstadt, Germany \label{addr3}
}

\date{Received: date / Accepted: date}

\maketitle

\begin{abstract}
We consider the Higgs potential in generalizations of the Standard Model. The possibility of the potential to develop two almost degenerate minima is explored. This would imply that QCD matter at two distinct sets of quark masses is relevant for astrophysics and cosmology. If in the exotic minimum the QCD matter ground state is electromagnetically neutral, dark matter may consist of QCD matter and antimatter in bubbles of the Higgs field. We predict an abundance of $\gamma$ rays in the few MeV region as messengers of dark matter regions in space. In addition the ratio of dark matter to normal matter is expected to show a time dependence.
\keywords{chiral symmetry \and Higgs potential \and dark QCD matter \and baryon asymmetry}
\PACS{12.38.-t \and 98.80Cq \and 12.60-i }
\end{abstract}


\section{Introduction}
Dark-matter studies receive considerable attention in fundamental research
(see e.g.~\cite{Bertone:2004pz,Rocha:2012jg,Tulin:2013teo,Petraki:2013wwa,Gordon:2013vta,Hui:2016ltb,Carr:2016drx,deSalas:2019pee,Montero-Camacho:2019jte,Grote:2019uvn,Bai:2019zcd,Savastano:2019zpr,Alvarez-Salazar:2019npi,Castellanos:2019kby,Ketov:2019qjw,DeNapoli:2019vcd}). 
Various scenarios proposed require new particles in extensions of the Standard Model (SM) (see e.g.~\cite{Bertone:2004pz,Hui:2016ltb}). 

The purpose of our Letter is to discuss a possible alternative of such scenarios based on exotic QCD matter.
In a recent work the authors presented a detailed study suggesting that QCD matter  depends crucially on the Higgs field \cite{Lutz:2014oxa,Lutz:2018cqo,Guo:2019nyp}. Within the SM the quark masses in QCD are proportional to the Higgs field. As a consequence, changing its ground state value does change the quark masses in QCD, however, in a manner that keeps all quark-mass ratios fixed. In \cite{Guo:2019nyp}
a possible first order transition along the Higgs field trajectory was discussed. It is compatible with current QCD lattice simulations of the baryon ground state masses, but should be scrutinized by further dedicated QCD lattice studies.

\begin{figure}
\centering
\includegraphics[width=0.44\textwidth]{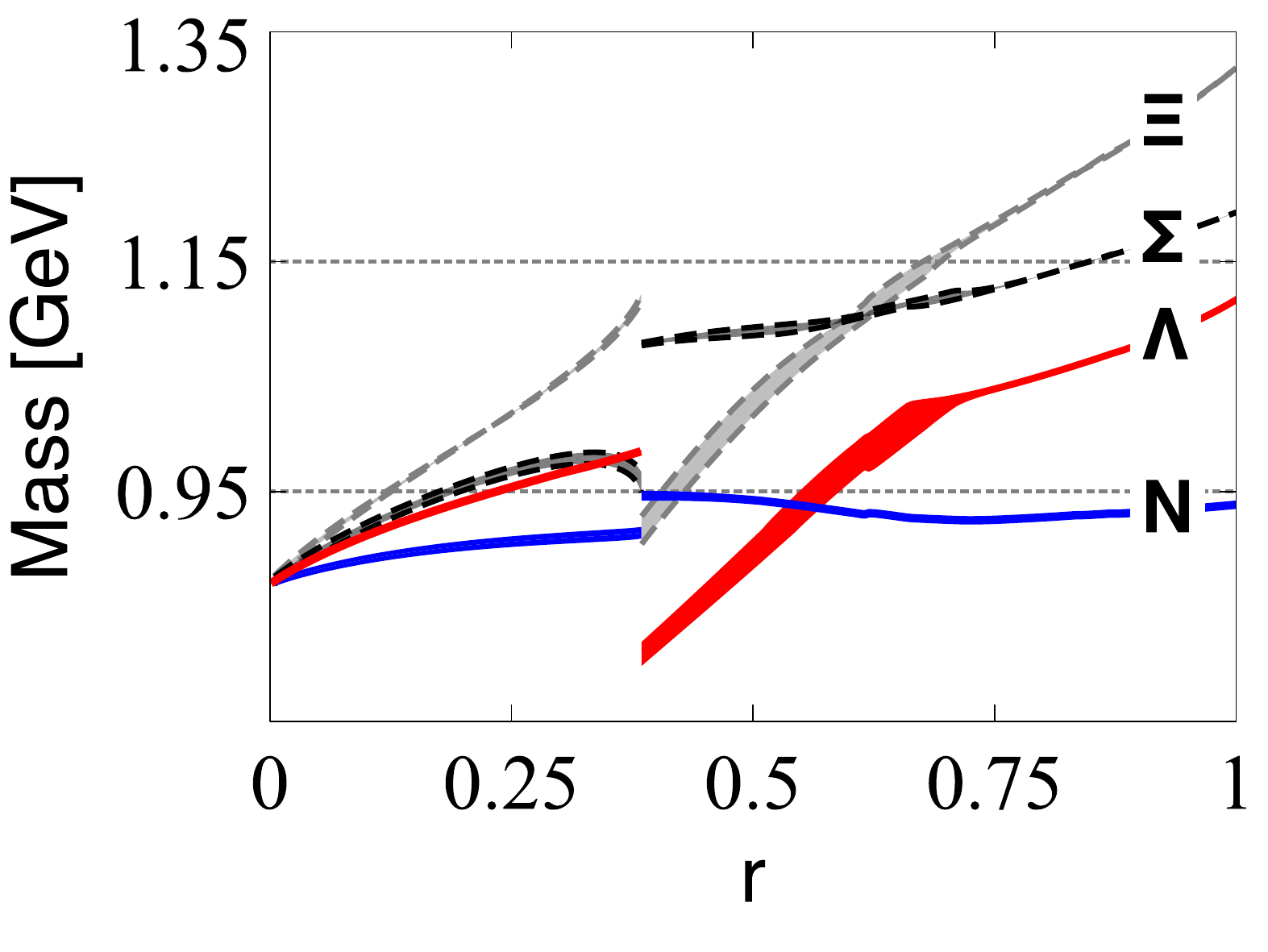}
\caption{Isospin averaged baryon masses as a function of $r = m_s/m^{\rm phys}_s$ along the Higgs trajectory with $2\,m_s/(m_u+m_d) = 26$ kept constant.  }
\label{fig-1}
\end{figure}

In Fig.\,\ref{fig-1} we show our prediction of the baryon masses along the Higgs trajectory \cite{Lutz:2014oxa,Lutz:2018cqo,Guo:2019nyp}. The bands in the plot provide an estimate of uncertainties based on our Fit 1 and Fit 2 scenarios as discussed in \cite{Guo:2019nyp}. 
At fixed ratio $m_s/m= 26$ the masses are plotted as functions of the strange quark mass. 
The key observation is that within a critical region of the Higgs field, baryonic matter and antimatter are composed from $\Lambda$ and $\bar \Lambda$ particles rather than from nucleons and anti-nucleons. This follows from the relation 
$M_\Lambda < M_N$, which holds at a specific range of the strange quark mass. 
We point out that our dark-matter scenario does not rely necessarily on a first order transition. Since $\Lambda$ particles are electromagnetically neutral such matter does not radiate and therefore appears dark. 
Since the Higgs sector of the SM drives a possible electroweak phase transition and underlies  baryogenesis models (see e.g~\cite{Noble:2007kk,Cao:2017oez,Kobakhidze:2015xlz,Reichert:2017puo}) it is important to explore exotic Higgs sector generalizations of the SM in more detail. Our dark matter scenario should not be confused with the quark nugget scenario proposed by Witten a long time ago \cite{Witten:1984rs}, in which 
dark matter would consist of deeply bound strange objects in the non-exotic Higgs phase.

\section{The Higgs potential}

We consider the Higgs sector of the SM \cite{Coleman:1973jx,Coleman:1977py,Martin:1997ns,Isidori:2001bm,Grojean:2004xa,Noble:2007kk,Buttazzo:2013uya,Spannowsky:2016ile,Reichert:2017puo,Gan:2017mcv,Jain:2017sqm,Braathen:2019pxr}. At tree-level the Higgs potential in the SM can be expressed in terms of two parameters only
\begin{eqnarray}
 V(H) = \frac{M_h^2}{2\,v^2}\,\Big( H^\dagger H -\frac{v^2}{2} \Big)^2\,,
 \label{def-SM}
\end{eqnarray}
with the complex doublet Higgs field $H$, the Higgs mass parameter $M_h \simeq 125.2$ GeV and the vacuum expectation value $v \simeq 246.2$ GeV of the Higgs field in its physical vacuum state \cite{Tanabashi:2018oca}. The value of $v$ plays a decisive role in the QCD part of the SM since all quark masses are proportional to $v$. 
In this work we are interested in the Higgs potential at $H^\dagger\,H \leq v^2$, where it is known that even loop corrections in the SM 
are sizeable (see e.g.~\cite{Coleman:1973jx,Martin:1997ns,Grojean:2004xa,Buttazzo:2013uya,Tamarit:2014dua,Spannowsky:2016ile,Hamada:2016iwx}). Since the Higgs potential will be affected in most extensions of the SM we follow here a phenomenological path where we explore the consequence of a fine-tuned potential with 
two degenerate minima. An effective field theory approach that implies two degenerate minima would require at least $(H^\dagger H)^3$ and 
$(H^\dagger H)^4$ operators. Consider the specific form
\begin{eqnarray}
&& V_{}(H) =  \frac{2\,M_h^2\,}{v^6\,(1-r^2 )^2}\,\Big( H^\dagger H -\frac{v^2}{2} \Big)^2\,\Big( H^\dagger H -\frac{v_a^2}{2} \Big)^2 \,,
\nonumber\\
&& {\rm and }\qquad r = v_a/v =m_s/m^{\rm phys}_s\,,
\label{def-model-1}
\end{eqnarray}
with $v_a$ the vacuum expectation value of the Higgs field at the exotic minimum. By construction, the model potential (\ref{def-model-1}) has two degenerate minima. At its physical one it recovers the empirical mass of the Higgs. The ratio $r=v_a/v$ determines the strange quark mass in the exotic minimum. 

The puzzle with (\ref{def-model-1}) is that it may be unnatural in the size of 
its dimension-full operators. However, we may recast the problem by considering loop corrections (see e.g~\cite{Coleman:1973jx,Martin:1997ns,Buttazzo:2013uya}). In the presence of multi-loop effects we may use the phenomenological ansatz
\begin{eqnarray}
&& V_{}(H) =  \frac{M_h^2\,}{2\,v^2\,[\log(\gamma + r^2 ) -\log (\gamma + 1) ]^2 }\,\Big( H^\dagger H -\frac{v^2}{2} \Big)^2\,
\nonumber\\
&& \qquad \;\,\times 
\Big( \log \big[\gamma + 2\,H^\dagger H/v^2 \big] -\log [\gamma + r^2] \Big)^2 \,,
\label{def-model-2}
\end{eqnarray}
where the particular form of the log term with the parameter $\gamma $ is taken from \cite{Spannowsky:2016ile}. 
There the value $\gamma =0.1$ is used. We note that the Higgs sector is  the 
least controlled part of the SM and therefore may be subject to significant model 
modifications.

\begin{figure}
\centering
\includegraphics[width=0.44\textwidth]{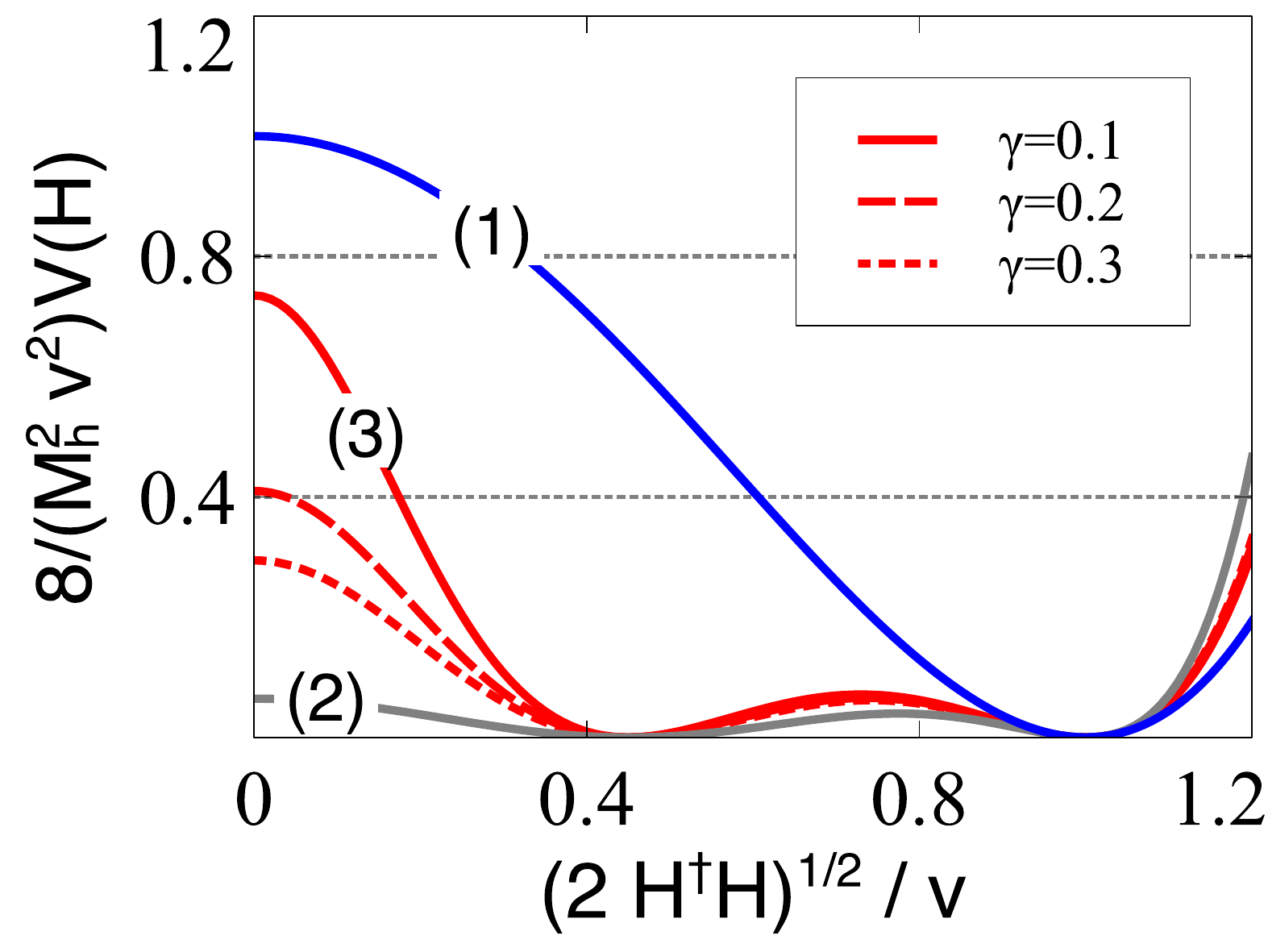}
\caption{A Higgs potential with two local minima as introduced in (\ref{def-model-1}) and (\ref{def-model-2}). 
It is compared with the tree-level potential of the SM (\ref{def-SM}).}
\label{fig-2}
\end{figure}

According to Fit 1 and Fit 2 we expect dark QCD matter in the range $ 0.39 < r < 0.57 $ and $ 0.39 < r < 0.54 $ respectively. The critical values are close to those  as derived in \cite{Guo:2019nyp} on the unphysical trajectory where $m_u+m_d$ is kept constant. 
In Fig.\,\ref{fig-2} we plot the effective potentials of (\ref{def-SM}-\ref{def-model-2}) 
as a function of $ \sqrt{2\,H^\dagger H } /v$ for the 
particular choice $r = 0.45$. The two degenerate minima are clearly visible for any of the three choices $\gamma = 0.1$, $\gamma = 0.2$ and $\gamma = 0.3$. With the parameter $\gamma$ we can efficiently dial the magnitude of the Higgs potential close to the origin. In the vicinity of the two local minima we find a rather mild dependence on the form of our 
parametrization. The polynomial ansatz (\ref{def-model-1}) or the log form (\ref{def-model-2}) lead to almost indistinguishable results. We emphasize that both models are compatible with 
empirical constraints on the Higgs potential as discussed in \cite{Cepeda:2019klc}. For instance at $r =0.45$ we extract from (\ref{def-model-1}) and (\ref{def-model-2}) the range $ 3.8 <  \kappa_\lambda < 6$ for the three Higgs coupling constant $\kappa_\lambda$. 
This is well compatible with the empirical 2-$\sigma $ interval $ -5.0 <\kappa_\lambda < 12.1$ from ATLAS \cite{Cepeda:2019klc}.  Our estimate excludes the SM value $\kappa_\lambda =1$.

One may object to such a fine-tuned Higgs potential. However, we wish to recall that there are ample cases in physics in which a system is driven by fine-tuned dynamical assumptions. In particular the SM itself has various fine-tuning issues already. 
At this stage of the development we would not worry too much. Rather, we  discuss in some detail the consequences of a possible dark QCD matter scenario.

Dark matter is believed to account for approximately 85$\%$ of the matter in the universe (see e.g.~\cite{deSwart:2017heh}). 
How would we arrive at such a  ratio in our scenario? We may assume here that the universe reheats above the electroweak scale after inflation. 
Alternative scenarios, where for instance the energy scale of inflation is around the electroweak scale (see e.g. \cite{Enqvist:2013kaa}), are not considered here.
Then the universe is baryon matter dominated already when the electrocweak symmetry is restored with $H = (0,0)$ (see e.g.~\cite{Jarlskog:1985ht,Shaposhnikov:1986jp,Gavela:1993ts,Hou:2011df,Spannowsky:2016ile}). 
As the temperature lowers further our effective Higgs potential should take over. Here it is important to estimate the size of thermal effects. In the limiting case of sufficiently large $T$  
they are proportional to $T^2\,V''(H)$ (see \cite{Dolan:1973qd}). Contrary to the SM, the finite temperature effect on the generalized Higgs potential is not monotonic in the Higgs field. From its 
functional form as modeled in Fig. 2 we expect the two local minima to merge to one global minimum. This is so since our two minima imply multiple sign changes in $V''(H)$. Thus, at this stage of the cosmic evolution, matter sits in a 'conventional' Higgs 
field condensate. Only as we further lower the temperature, the additional exotic minimum will turn visible. We do not see any strong hint that in the electroweak era of the cosmic evolution 
any seed of dark matter is formed.

\begin{figure}
\centering
\includegraphics[width=0.44\textwidth]{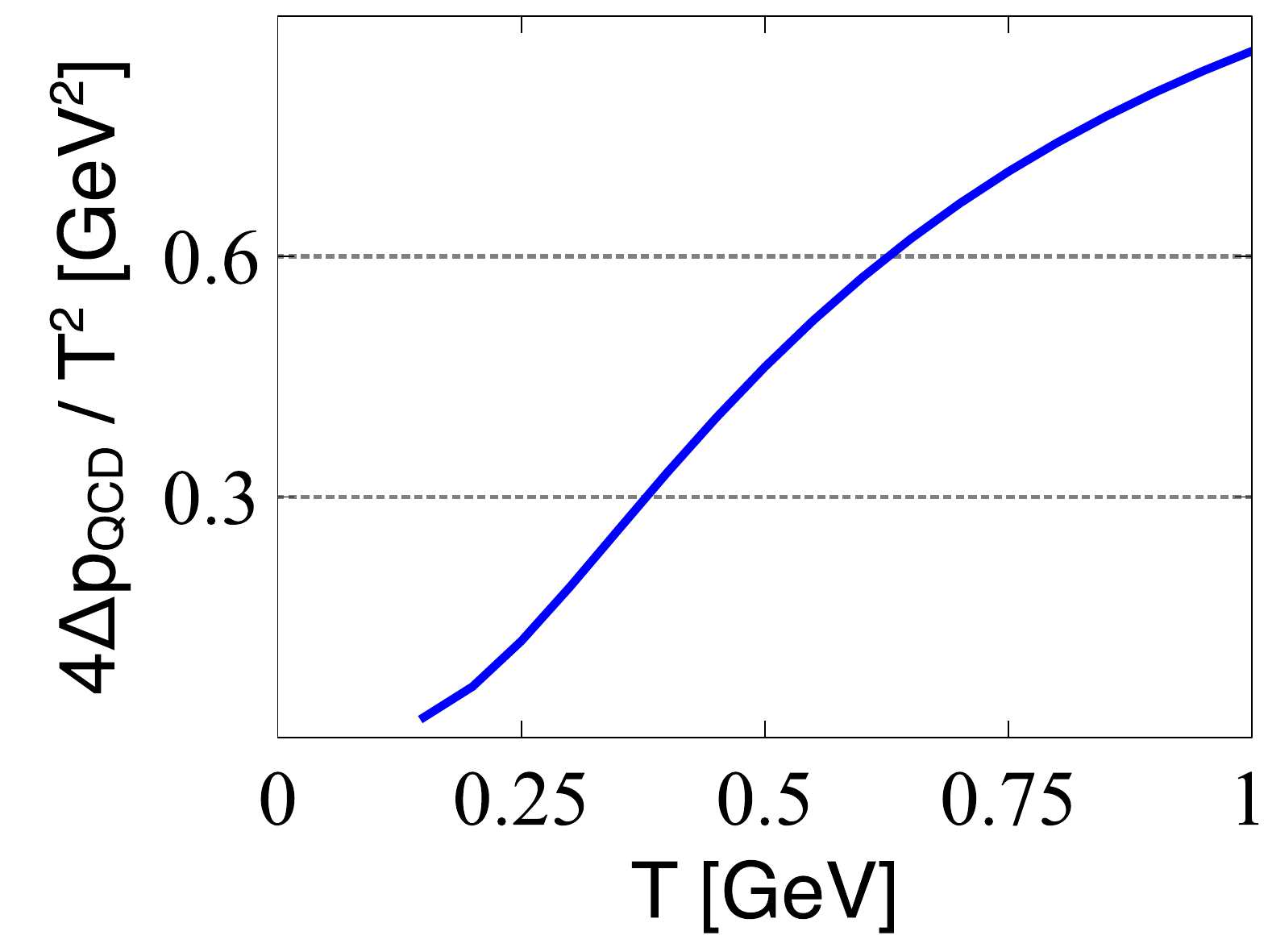}
\caption{Difference of the pressure densities, $\Delta p_{QCD}$, in the two Higgs phases as a function of temperature. We use $m_{u,d} = 4 $ MeV, $m_s = 26\,m_{u,d}$ and $m_c= 12 \,m_s$ with r = 0.45. }
\label{fig-2b}
\end{figure}

At the QCD scale $T \sim 1$ GeV thermal effects from quarks and gluons will dominate the temperature effects in the Higgs potential. Here the competetion of the two possible Higgs 
phases is more intricate. Despite our assumption on the almost degeneracy of the two Higgs phases at zero temperature, there will be a significant asymmetry from the dynamics of 
quarks and gluons. It is safe to assume a vanishing baryon chemical potential and consider the difference of the pressure densities in the two available local minima. The exotic 
Higgs phase phase wins here, since it is characterized by quark masses that are about twice as small as those in the normal phase. This is illustrated in Fig. \ref{fig-2b}, where 
the difference in the pressure densities of the two Higgs phases is shown as a function of the temperature $T$. At $T= 1$ GeV that difference of about $\Delta p_{QCD}  \simeq 28$ GeV/fm$^3$ 
is dominated by the presence of the heavy quark. If we ignored the charm quark the corresponding value would be $\Delta p_{QCD} \simeq 0.28 $ GeV/fm$^3$.
The computation follows from a non-interaction fermi gas ansatz for the quark-gluon-plasma phase of QCD, which may serve as a rough estimate. 
Note that, the contributions from gluon degrees of freedom cancel out in the considered difference, at least approximatly \cite{Fogaca:2010mf,Sanches:2014gfa}. 
Thus the results are determined by the quark masses in the two phases and the temperature only. 
In the limit of large temperatures we find,  
$\Delta p^{(q)}_{QCD} \simeq (1-r^2)\,4\,m_q^2\,T^2$, for a given quark flavor of mass $m_q$. 
In the non-exotic phase we used isospin averaged values of $m = 4 $ MeV 
for the up and down quark masses, $m_s = 26\,m$ for the strange quark mass and $m_c= 12 \,m_s$ for the charm quark mass. The corresponding masses in the exotic phase are 
implied by the ratio parameter $r = 0.45$. Altogether, the exotic phase wins here and therefore exotic Higgs bubbles will start to form.

Eventually the formation of such exotic bubbles will stop as the temperature is further reduced. We expect this to happen after QCD changed its degrees of freedom from quarks 
and gluons to hadrons at temperatures $T < 150$ MeV. Thus, altogether we suggest that dark baryonic matter is formed via a first-order phase 
transition taking place in the quark-gluon plasma era of the cosmic evolution. To reach the target of 15$\%$ ordinary to dark matter ratio appears realistic, however, 
requires detailed knowledge of QCD dynamics during that phase transition. A significant estimate of that ratio requires a quantitative study of the evolution process. 
That is beyond the scope of the current work.

\begin{figure}
\centering
\includegraphics[width=0.44\textwidth]{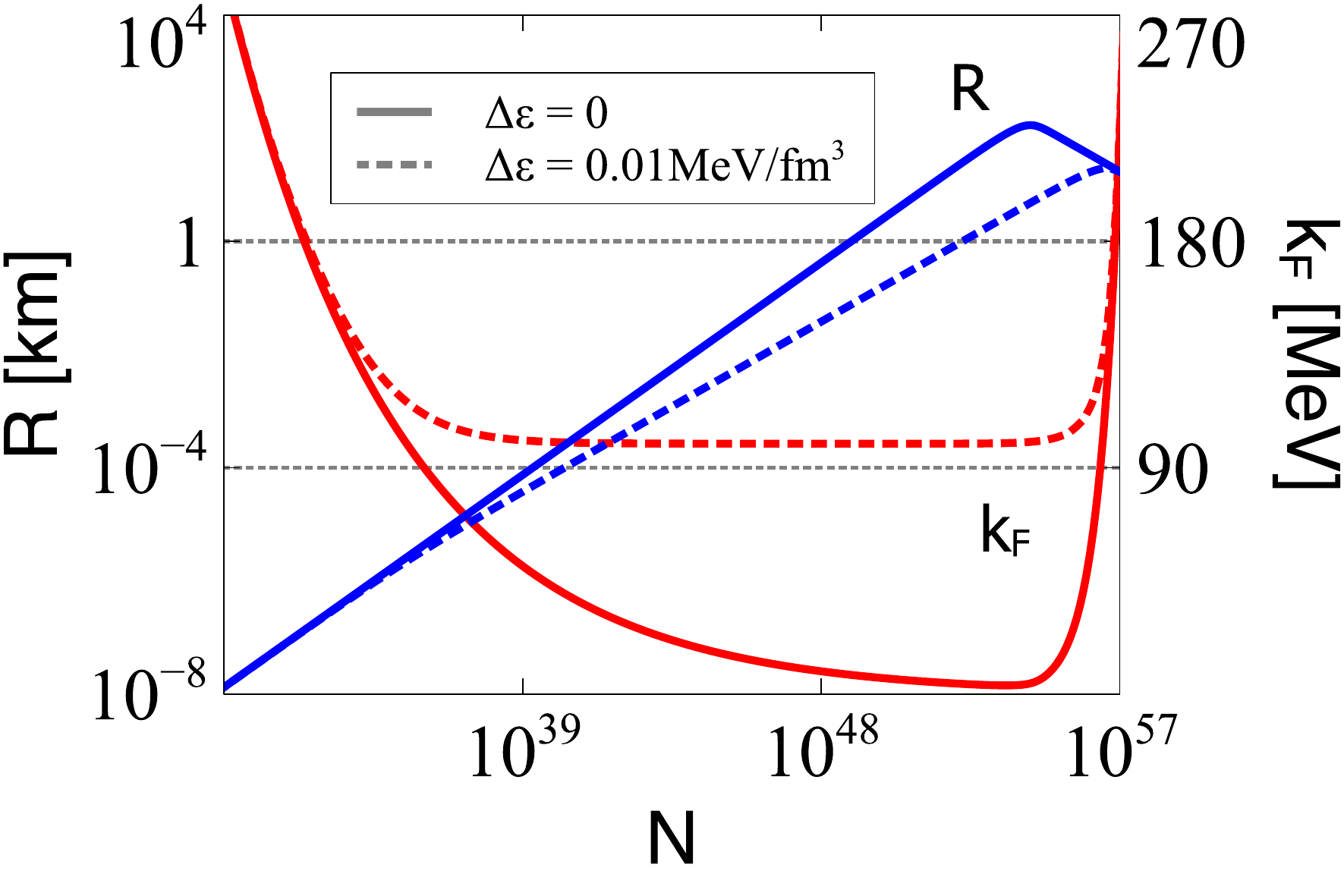}
\caption{The radius and Fermi momentum as a function of $N$ at fixed $\Delta \epsilon $.  Note that $N_{\rm Sun} \simeq  M_{\rm Sun }/ M_{\rm Nucleon} \simeq 1.2 \times 10^{57}$. }
\label{fig-3}
\end{figure}

If the exotic minimum in the Higgs potential is slightly metastable at zero temperature, we expect a scenario where the vacuum shows bubbles with dark QCD matter inside, but normal QCD matter outside. Inside the bubbles the matter or antimatter ground states consist of $\Lambda$ or $\bar \Lambda$ particles, however with exotic properties as shown by Fig.\,\ref{fig-1}. 

Let us explore 
the stability of a possible Higgs bubble. Since the boundary of such a bubble 
stores a significant amount of energy, there is a tendency that such a bubble shrinks or even 
collapses. From Fig.\,\ref{fig-2} we estimate the energy density
\begin{eqnarray}
 \epsilon_{\rm Higgs} \simeq 1.1 \times 10^9 \,{\rm GeV}/{\rm fm}^3
\end{eqnarray}
from the Higgs potential taking in-between its two minima. We now assume a Higgs bubble with spherical geometry characterized by a radius $R$ and a surface thickness $d$. That implies
the total surface energy
\begin{eqnarray}
 E_{\rm surface} = 4\,\pi\, d\,R^2 \,\epsilon_{\rm Higgs}  + 4\,\pi R^2\,(\Delta v)^2/d \,,
 \label{res-surface}
\end{eqnarray}
where the second term in (\ref{res-surface}) follows from the kinetic term of the Higgs field. From Fig.\,\ref{fig-2} we can read off the change of the Higgs field across the surface with  $ (\Delta v)^2 \simeq v^2/8$. 
We estimate the bulk energy by a free-Fermi gas approximation 

\begin{eqnarray}
&& E_{\rm bulk} = 
\Big(M_{\Lambda, in}  + \frac{3}{10}\,k_F^2/M_{\Lambda, in} \Big) \,N 
 +\,\frac{4\,\pi\,R^3}{3}\,\Delta \epsilon\nonumber\\
&&\qquad\qquad
-\frac{3}{5}\,\frac{M_{\Lambda,in }^2}{R}\, G\,N^2\,, \qquad \qquad 
\nonumber\\
&& N =  \frac{4\,(R\,k_F)^3}{9\,\pi}\, ,
\label{res-bulk}
\end{eqnarray}
with the gravitational constant $G \simeq 6.709\times 10^{-39}\, {\rm GeV}^{-2}$ and $N$ the total number of $\Lambda$'s in the Higgs bubble. Their Fermi momentum is denoted by $k_F$ with $\rho=k_F^3/(3 \,\pi^2)$, where $\rho$ specifies the dark-matter density in the bubble.
We parameterize a supposedly small difference in the vacuum energy densities 
at the two Higgs minima by $\Delta \epsilon > 0$, where we assume 
the dark-matter vacuum to be slightly disfavored.

A Higgs bubble can be stable provided that it encloses a sufficient amount of dark matter. 
We can make this more quantitative by a minimization of its energy $ E= E_{\rm bulk} +E_{\rm surface} $ with respect to the surface thickness $d $ and the radius $R$ at a fixed value of the 
total number of $\Lambda$'s in the bubble. From this we find the two relations,
\begin{eqnarray}
&&  d \simeq \sqrt{ (\Delta v)^2/\epsilon_{\rm Higgs}} \simeq 6 \times 10^{-3} \,{\rm fm}  \,,\qquad 
\nonumber\\
&& \frac{1}{R} + \frac{R^2\,G\,k_F^6\,M^2_{\Lambda, in}}{135\,\pi^2\,d\,\epsilon_{\rm Higgs}}=
  \frac{\Delta \epsilon}{4\,d\,\epsilon_{\rm Higgs}}\,\nonumber\\
&& 
\qquad\qquad\qquad\qquad\qquad \times\Bigg( \frac{k^5_F}{15\,\pi^2\,M_{\Lambda, in} \,\Delta\epsilon} - 1\Bigg)\,.
\label{res-dR}
\end{eqnarray}
This implies that at given $k_F$ it follows that  $\Delta \epsilon $ must be smaller than a critical value,
\begin{eqnarray}
 \Delta \epsilon <  \Delta \epsilon_{crit} =\frac{k_F^5}{15\,\pi^2\,M_{\Lambda, in} }\,,
\end{eqnarray}
as to keep the dark-matter bubble stable. We checked that all second derivatives are 
positive so that with (\ref{res-dR}) we have at least a local minimum of the dark matter system (\ref{res-surface}, \ref{res-bulk}). 
In Fig. \ref{fig-3} we show the radius, $R$, and Fermi momentum, $k_F$, of the Higgs bubble as a function of $N$ at various fixed values of $\Delta \epsilon$. 
Within the range $ 10^{31} < N < 10^{57}$ the value of $k_F < 250 $ MeV is small enough to justify our free-Fermi gas approximation. We expect our results to hold at the qualitative level.  

\begin{figure}
\centering
\includegraphics[width=0.44\textwidth]{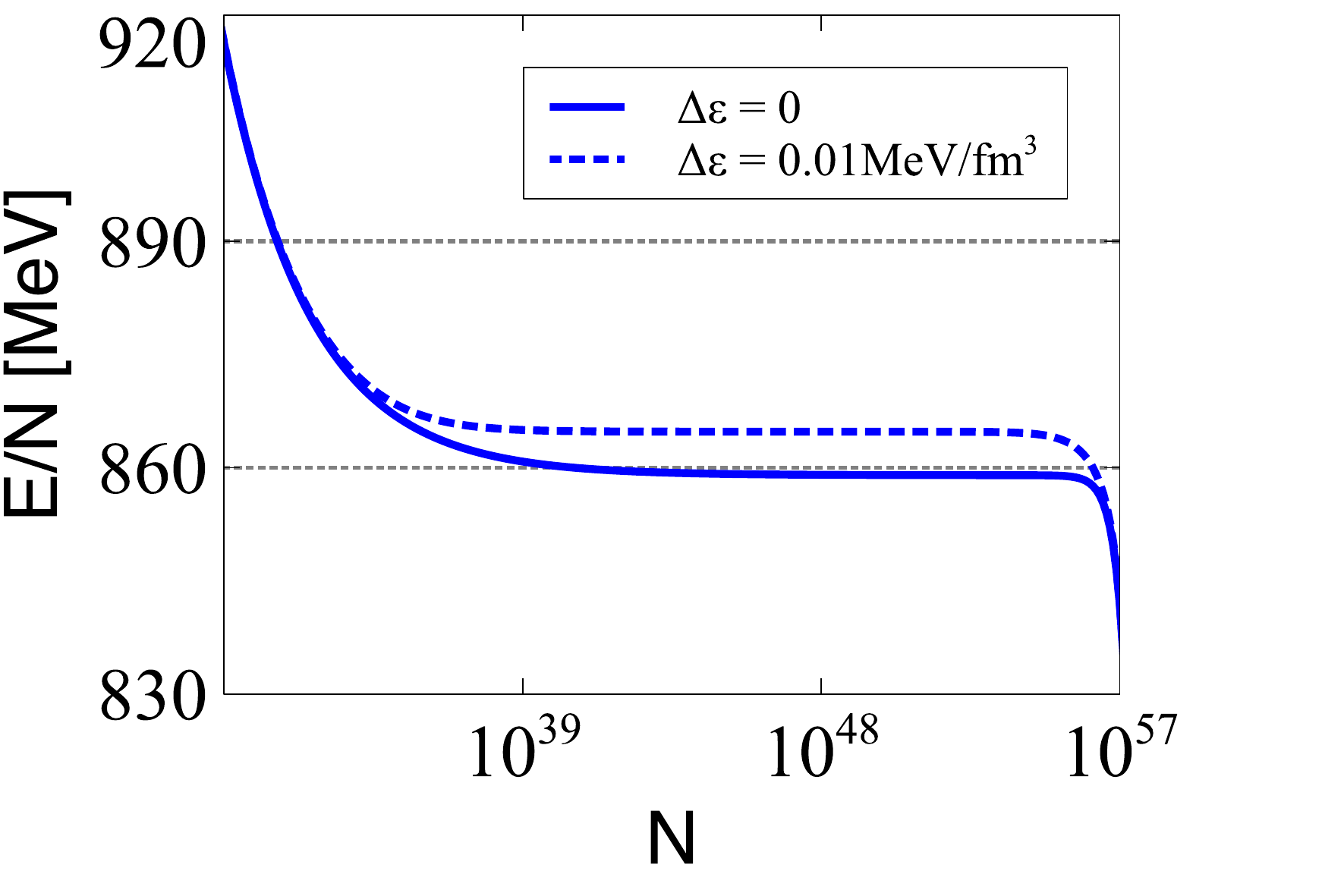}
\caption{The energy per particle $E_\Lambda/N$ as a function of $N$ at fixed $\Delta \epsilon $.  Note that $N_{\rm Sun} \simeq  M_{\rm Sun }/ M_{\rm Nucleon} \simeq 1.2 \times 10^{57}$. }
\label{fig-4}
\end{figure}

It is left to check whether such a dark matter Higgs bubble is stable with respect to 
a decay into a more conventional object consisting out of normal baryonic matter. 
A useful quantity to consider is the energy per particle in the bubble, $(E_{\rm surface}+E_{\rm bulk} )/N$,
with $d$ and $R $ as given in (\ref{res-dR}). 
In Fig. \ref{fig-4} we show such a dependence at various fixed values of $\Delta \epsilon$. 
The energy per particle is significantly smaller than the free nucleon mass and therefore, at least in the region $ 10^{31} < N < 10^{57}$, where the effects from gravity are not dominating the system yet, there is no phase-space available for such a decay.

As an example consider $N = M_{\rm sun}/ M_N \simeq 1.2 \times 10^{57}$ for which it follows $R \simeq 16$ km and $k_F \simeq 250$ MeV at $\Delta \epsilon = 0$. The particle density 
with $\rho \simeq 0.07\, $fm$^{-3}$ would be sufficiently small as to justify the application of the free-Fermi gas approximation. As we further increase $N > N_{\rm Sun}$ the dark matter will turn more dense and  will be sensitive to the equation of state of the $\Lambda$ particles in the Higgs bubble. It would be important to establish the latter from QCD and to derive mass limits for the dark matter bubbles. 
Here the  Tolman-Oppenheimer-Volkoff equation has to be supplemented by a suitable boundary condition at the Higgs bubble surface. This may open the possibility for the existence  of massive compact objects, with properties distinct to those of neutron stars and/or black holes of unconventionally small masses \cite{Thompson637}. 

In a meson-exchange phenomenology, the repulsive omega-meson exchange process is expected to dominate the short-range interaction of $\Lambda$ particles in the Higgs bubble. Due to the approximate isospin conservation, pion- and rho-meson exchange processes are suppressed. It remain the eta-meson exchange and the two-pion exchange contributions, which may bring in some weak intermediate-range attractive forces \cite{Sasaki:2006cx,Polinder:2007mp,Bauer:2015ega}.

While there appears to be a rather weak net attraction at the physical point \cite{Bauer:2015ega,Haidenbauer:2015zqb,Sasaki:2015ifa,Sasaki:2018mzh} available studies suggest a sizeable quark-mass dependence thereof \cite{Sasaki:2006cx,Beane:2011zpa,Sasaki:2015ifa,Sasaki:2018mzh}. We conclude that at the exotic Higgs minimum, that comes at much smaller up and down quark masses, there is little evidence to expect this weak attraction to survive. Whether and how massive dark-matter clusters form depends on the subtle balance of the gravitational force and the short-range strong interactions in the Higgs bubble.

We conclude that in any case the typical dark-matter density in a Higgs bubble should be significantly larger than the density of a cold interstellar medium, which is characterized by a baryon-number density smaller than about $10^6/ {\rm cm}^3$.
In this context we discuss the so-called Bullet Cluster \cite{Clowe_2004,Markevitch_2004}. While the radial velocity distributions of stars inside a galaxy or data  on  gravitational lensing effects (see e.g. \cite{Newman:2009qm,deSwart:2017heh,Helmi:2018,deSalas:2019pee,Relatores:2019ceq}) put constraints on the dark matter 
distributions in and outside galaxies, more significant information on the possible nature of 
dark matter is set by the observation of collisions of galaxy clusters \cite{Clowe_2004,Markevitch_2004}. 
It is found that in such a collision there is no direct hint pointing at any sizeable interaction of dark matter with ordinary matter \cite{Markevitch_2004}. 
In this context we have to discuss how a Higgs bubble interacts with protons from the intergalactic hot gas. The relative velocity of the two colliding galaxies in  \cite{Clowe_2004} is of the order of 4500 km/s. An intergalactic gas of temperature $T \simeq 6$ keV implies a typical proton velocity of about 1300 km/s. Thus most of the protons from the gas do not have sufficient kinetic energy to invade the bubble. In turn there will be no strong interaction effects visible. Second we need to consider the case where Higgs bubbles from the two galaxies collide. The chance that this happens depends on the typical size of such bubbles, which are not well constrained at this stage. They depend on the details of the Higgs potential, in particular the size of $\Delta \epsilon$ term, and a cosmological model. The smaller the typical size of the Higgs bubbles, the smaller the likelihood that such a process turns relevant in a galaxy merger event. Even if two bubbles start to overlap, we would expect that the two bubbles merge into a larger one, since this reduces the energy stored in their surface. The residual interaction of the Lambda particles with kinetic energies of at most a few MeV should be dominated by elastic processes. In turn we do not see any reason to expect a strong visible effect of the dark matter component in such a galaxy collision event.  

Last, we  turn to a most interesting process where a sufficiently energetic cosmic proton tries to enter a dark-matter region in space with relative velocity, $v_p$. Note that, depending on the energy such a proton may be even trapped inside the dark-matter bubble and therefore 
the ratio of dark matter to normal matter is expected to show a time dependence in our dark-matter scenario \cite{Genzel:2017jgd}. 
According to Fig.\,\ref{fig-1} the nucleon mass inside the bubble is only up 10 MeV $= \Delta M_N$ larger than its mass outside the bubble. Thus on the way into the bubble the nucleon has to either transfer momentum to the Higgs bubble and/or radiate photons. 
Such a Bremsstrahlung spectrum should be limited to $\gamma$ rays with energies less than that 10 MeV. Here a crucial parameter is the acceleration, $a \simeq \,(c^2/\gamma^2_p)\,\Delta M_N/(d\,M_N ) \simeq 8\times 10^{23}\,(c/\gamma^2_p)/s$, of the proton across the Higgs bubble surface, since its total radiation power is proportional to $a^2 \,\gamma_p^4$ with $\gamma_p =(1-v_p^2/c^2)^{-1/2}$. 
To this extent our Higgs bubbles glim with a characteristic spectrum which depends on the details of the Higgs potential. 

So far we made a rough estimate for the flux of sub MeV photons. To be explicit, we assumed a dark matter bubble 
of radius 16 km moving  with 4500 km$/$sec through an inter galactic gas of 6 keV temperature inside the bullet cluster. For this, the integrated X-ray flux arriving at a detector on the earth is $ 10^{-48}$ Watt$/$cm$^2$. It is instructive to confront this value with the integrated flux of $5.6\times  10^{-19}$ Watt$/$cm$^2$  for X-rays with (0.1-2.4) keV 
from the Bullet Cluster \cite{Clowe_2004,Markevitch_2004}. Our corresponding estimate for integrated X-ray flux from 
dark matter bubbles is less than $ 10^{-36}$ Watt$/$cm$^2$, i.e. down by 17 orders of magnitude. A similar value we predict from the photon emission of a corresponding single dark matter bubble in the Milky Way, for which we estimate the rate to be smaller than $ 10^{-38}$ Watt$/$cm$^2$.

It is not very likely that such dark matter photons can be detected with satellite-based detectors like e-Astrogram or AMEGO \cite{Hill:2018trh,mcenery2019allsky,Oakes:2019ywx}. In particular, we note that so far there is gamma ray data in the (1-10) MeV region available with quite large uncertainties only.

\section{Summary and conclusions}

We constructed a phenomenological Higgs potential with two degenerate local minima. It was argued that such a generalization of the SM may lead to dark QCD matter that lives in bubbles of the Higgs field, with normal QCD matter outside and dark QCD matter inside. Within the bubbles we expect exotic $\Lambda $ and $\bar \Lambda$ particles, that are formed by QCD at unconventionally small up, down and  strange quark masses. We predict an abundance of $\gamma$ rays in the few MeV region as messengers of dark matter regions in space.  
In addition the ratio of dark matter to normal matter is expected to show a time dependence.

It would be interesting to further scrutinize the dark QCD matter scenario proposed here. With current QCD lattice techniques it is possible to substantiate or rule out such a scenario by further studies of the strange quark-mass dependence of the nucleon and $\Lambda$ baryon  masses. It would be important to establish a more fundamental framework in which such an exotic Higgs potential is implied.

\begin{acknowledgements}
Bengt Friman, Hans Feldmeier, Gregor Kasieczka, Evgeny Kolomeitsev, Thomas Mannel, Guy Moore, 
Walter Sch\"on, Madeleine Soyeur, Christian Sturm and David Urner are acknowledged for stimulating discussions. Y. H. received partial support from Suranaree University of Technology, Office of the Higher Education Commission under NRU project of Thailand (SUT-COE: High Energy Physics and Astrophysics) and SUT-CHE-NRU (Grant No. FtR.11/2561).
\end{acknowledgements}


\bibliographystyle{spphys}
\bibliography{literature}

\end{document}